\documentclass[prl,twocolumn,showpacs,preprintnumbers,amsmath,amssymb]{revtex4}
\usepackage{graphicx,lscape}
\usepackage{dcolumn}
\usepackage{bm}
\begin{document}
\renewcommand{\baselinestretch}{1}

\title{Optical properties of photonic crystal slabs with asymmetrical unit cell}

\author{
N.~A.~Gippius,$^{1,2}$ S.~G.~Tikhodeev,$^{1,2}$
and T.Ishihara$^{2}$ }

\affiliation{
$^1$ General Physics Institute RAS, Vavilova 38, Moscow 119991, Russia \\
$^2$ EEL-FRS RIKEN, 2-1Hirosawa, Wako, Saitama, 351-0198, Japan }

\begin{abstract}
Using the unitarity and reciprocity properties of the scattering matrix,
we analyse the symmetry and resonant optical properties of the
photonic crystal
slabs (PCS) with complicated unit cell.
We show that
the reflectivity is not changed upon the 180$^\circ$-rotation of the sample
around the normal axis,
even in PCS with asymmetrical unit cell. Whereas the transmissivity
becomes asymmetrical if the diffraction  or absorption are present.
The PCS reflectivity
peaks to unity near the quasiguided mode resonance for normal light
incidence in the absence of diffraction,
depolarisation, and absorptive losses. For the oblique incidence
the full reflectivity is reached only in symmetrical PCS.
\end{abstract}

\pacs{
42.70.Qs, 42.25.Bs}
\date{Version of 27 February 2004}


\maketitle

The physics of photonic crystal slabs (PCS)~\cite{Maradudin93,Labilloy97,Astratov99}
receives much attention in recent years because of many interesting
possibilities they open to control the light interaction with
matter.
The tendency is toward the PCS  sophistication:
nanostructured metals or semiconductors
are included, and the unit
cell geometry, using  modern technology, becomes more complicated.
As a result, new tools for photonic engineering become available.
A well known example is the extraordinary optical transmission through sub-wavelength hole
arrays in metal films~\cite{Ebbesen98,Ghaemi98}.
Another promising example are polaritonic crystal
slabs with nanostructured semiconductors~\cite{Fujita98,Yablonskii01,Shimada02}.
The physics behind is the coupling between photonic and different
electronic resonances in such structures.
They manifest
themselves via pronounced
resonant Wood's anomalies~\cite{Wood902,Fano41,Hessel65} in the optical spectra,
due to the excitation of quasiguided~\cite{Labilloy97,Astratov99,Paddon00,Ochiai01-01,Fan02,Tikhodeev02}
or surface plasmon~\cite{Ritchie68,Ebbesen98,Schroeter99,Dykhne03} or
both~\cite{Christ03} modes.
On the other hand, making complicated unit cells, e.g.,
characterized by a lack of 180$^o$-rotational symmetry in the PCS plane~\cite{Schroeter99,Fujita00}
adds new ways to control the interaction with light.

Thus, the understanding of the symmetry and resonance properties
of the optical response in PCS with a complicated unit cell becomes important,
also for their optical characterization.
Measuring reflection  from asymmetric structure appears to be not a promising
method for the PCS
optical characterization, and a question arises, which optical properties are more sensitive to
the PCS structure?
Meanwhile, notwithstanding a long history of the investigations
(see, e.g., in~\cite{Hessel65,ERNO}),
starting actually from the optical gratings, which can be understood as
one-dimensional (1D) PCS, their
properties look sometime amazing and even contradictory.

One example is the nontrivial symmetry properties of
reflection and transmission from asymmetric PCS.
Namely, it was demonstrated~\cite{Schroeter99} experimentally and numerically
that the reflectivity
from  180$^o$-rotationally non-invariant (in the PCS plane) metal
gratings on a dielectric substrate is always symmetric, whereas the transmission is not.
However, the transmission appears to be independent on the side of the illumination,
whether it is from the air or from the substrate (see also in Refs.~\onlinecite{Ebbesen98,Ghaemi98}).
The authors~\cite{Schroeter99} find this \textit{astonishing}, because
the calculated fields distributions inside the PCS appear to be very different for the
illumination from different sides.

Another example is the so called anomalous full  reflection in zeroth diffraction order
in transparent PCS, when reflectivity peaks to unity resonantly~\cite{Hessel65}.
It is established that excitation of surface or quasiguided
modes in PCS is responsible for this resonant behavior. Such resonances, being Fano-type
discrete states in the continuum background, are characterized by a finite frequency linewidth due
to radiative losses even in transparent materials. The existence of radiative losses seems to exclude
a possibility of full reflectivity, however, all the
models~\cite{Hessel65,Golubenko85,Mashev85,Wang93,Peng96,Fan02} show
this effect in transparent PCS.
Moreover, it appears that very different physical models, with only one common thing, the existence
of \textit{any} resonance, predict a qualitatively similar behavior of the PCS optical response,
see, e.g., discussions in~\cite{Porto99,Garcia-Vidal02,Fan02,Dykhne03}.

The most general reasons of such behavior  are hidden in the most general
properties of the scattering matrix operator of the PCS. These general properties are actually well known
for many years, but, to the best of our knowledge,
 their consequences for the photonic crystal slabs with complicated
unit cell have not been yet investigated in detail.
In this Letter, we recall the most general properties of the S-matrix of the transparent
arbitrary PCS and use them to analyse the symmetry and resonance properties of the
optical response, including
the conditions of the anomalous full reflectivity.

Starting from a general scattering matrix operator of a planar periodic system
made of transparent materials,
we note that many scattering channels are closed, e.g., the backscattering for
inclined light incidence, as well as all non-Bragg scattering channels. In order
to keep all nonzero general S-matrix elements, it is convenient to
define the scattering matrix operator
as
a \textit{unitary} $2N\times 2N$-dimensional  matrix $S_u$,~
$S_u^\dag S_u = S_uS_u^\dag =1$, coupling the amplitudes of incoming and outgoing
harmonics (main
and diffracted). Here  $N = 2 + N_\mathrm{open,air} + N_\mathrm{open,sub}$,
the augend 2 stands for the main harmonic (in air and substrate), $N_\mathrm{open,air}$ and
$N_\mathrm{open,sub}$ are the numbers of open diffraction
orders (into the air and substrate claddings). Due to two polarisation states per each
harmonic (for example, $s$ and $p$, or $\sigma^+$ and $\sigma^-$),
there are $2N$ incoming as well as outgoing amplitudes (we assume isotropic air and substrate
claddings).
$S_u= S_u(\omega,\mathbf{k}_\parallel)$ is a function of the real incoming
photon frequency $\omega$ and in-plane wavevector
$\mathbf{k}_\parallel = (k_x,k_y) =
\frac{\omega}{c} \sin \vartheta (\cos \varphi, \sin \varphi)$,
where $\vartheta, \varphi$ are the azimuthal and polar angles of light incidence.
The wavevectors  of the diffracted photons are
\begin{eqnarray}\label{kG}
\mathbf{k}_{\parallel\mathbf{G}} &=& (k_{x\mathbf{G}},k_{y\mathbf{G}}),\, k_{x(y)\mathbf{G}}= k_{x(y)} + G_{x(y)},
\\ \label{kGz}
k_{z\mathbf{G}} &=&
\pm \sqrt{\omega^2 \varepsilon/ c^2 - k_{x\mathbf{G}}^2- k_{y\mathbf{G}}^2},
\end{eqnarray}
where  $\mathbf{G} = (G_x,G_y)$ is the reciprocal 2d PCS lattice,
and $\varepsilon = 1$ or $\varepsilon_\mathrm{sub}$,
depending on the PCS side; $k_{z\mathbf{G}}$ is real for open diffraction channels.
Such unitary scattering matrix $S_u$ can be constructed
from the infinitely-dimensional ``large'' scattering
matrix $\mathbb{S}$ accounting for the near-field coupling between all the propagating
and evanescent harmonics and defined, e.g., in Ref.~\onlinecite{Tikhodeev02}.
$\mathbb{S}$ has to be reduced~\cite{ERNO} to a ``small'' $2N\times 2N$-dimensional  matrix $S$
for propagating harmonics only
and transferred into any energy-flow-orthogonal
basis (e.g., of $sp$ or $\sigma^\pm$ polarisations),
we do not dwell into details here.

\begin{figure}[t]
\includegraphics[width=0.95\linewidth]{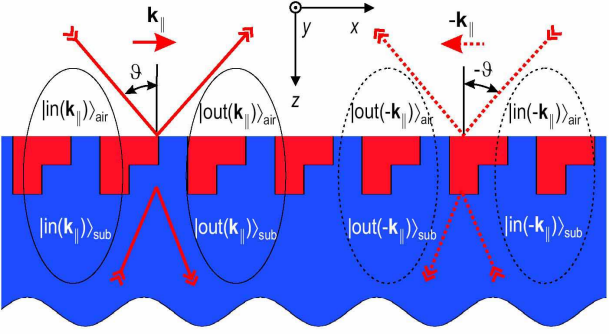} 
\caption{
Geometry of light reflection/transmission from PCS. Solid and dashed arrows correspond
to a direct and time reversed process. The main harmonics are shown only.
}
\label{fig:1a}
\end{figure}

The symmetry of the system with respect to the time reversal in the case
of  PCS made of non-gyrotropic transparent materials [compare with
Ref.~\onlinecite{Figotin01}], means that for
any solution  $E(\omega,\mathbf{k}_\parallel),H(\omega,\mathbf{k}_\parallel)$
of Maxwell equations for electric and magnetic fields,
$E^*(\omega,-\mathbf{k}_\parallel),-H^*(\omega,-\mathbf{k}_\parallel)$ remains
a valid solution. This property, as well known~\cite{LLQM}, leads to the
\textit{reciprocity} between the input and output channels.
In the definition of the unitary $S_u$
it makes sense to fix the ordering of the channels
in such a way that the input channels in $S_u(\omega,\mathbf{k}_\parallel)$ transfer into the output ones
in $S_u(\omega,-\mathbf{k}_\parallel)$ and vice versa, upon time reversal, see a scheme in Fig.\ref{fig:1a}.
In the case of $s,p$-polarisations basis, this  can be written as~\cite{endnote31}
\begin{equation}\label{Reciprocity}
|\textrm{in}\rangle \equiv
|\textrm{in}(\omega,\mathbf{k}_\parallel)\rangle =
|\textrm{out}(\omega,-\mathbf{k}_\parallel)^*\rangle
\equiv|\widetilde{\textrm{out}}\rangle.
\end{equation}
Note the important complex conjugates in  Eq.~(\ref{Reciprocity}). Then,
the reciprocality of the unitary $S_u$ matrix means that~\cite{endnote32}
\begin{equation}\label{resipr}
S_u(\omega,\mathbf{k}_\parallel) =  S_u^\intercal(\omega,-\mathbf{k}_\parallel),
\end{equation}
where ${}^\intercal$ stands for the matrix transpose. The most general form of $S_u$ allowing for the time reversal
can be then written as
\begin{equation}\label{SVD}
S_u = \sum_{j = 1}^{2N} \mathrm{e}^{i\beta_j}|\mathrm{out}\rangle_{j\,j}\langle \mathrm{in}| =
\sum_{j = 1}^{2N} \mathrm{e}^{i\beta_j}|\mathrm{out}\rangle_{j\,j}\langle \widetilde{\mathrm{out}|},
\end{equation}
where the input (output) orthogonal bases $|\mathrm{in}\rangle_j$ ($|\mathrm{out}\rangle_j$)
and the scattering phases $\beta_j,\, j = 1, \ldots 2N$ are functions
of $(\omega, \mathbf{k_\|})$ which are characteristic for the given PCS. The phases are
actually even functions of $\mathbf{k_\|}$,
$\beta_j(\omega,\mathbf{k_\|})=\beta_j(\omega,-\mathbf{k_\|})$.
Equation~(\ref{SVD}) means the existence of a special ``diagonal'' input and output
basis sets, which do not mix up in the process of scattering.

\begin{figure*}[t]
\includegraphics[width=0.34\linewidth]{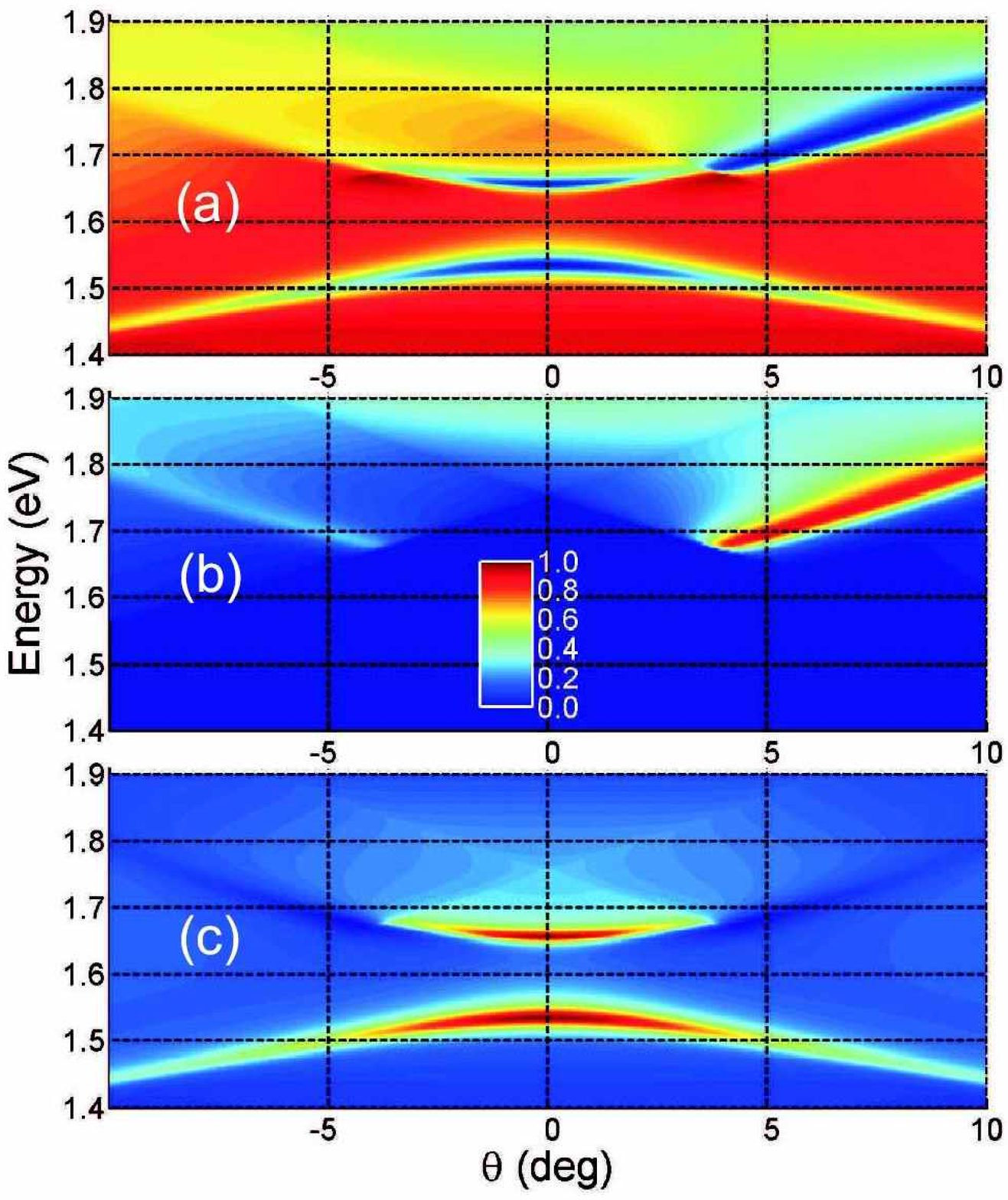}
\includegraphics[width=0.6\linewidth]{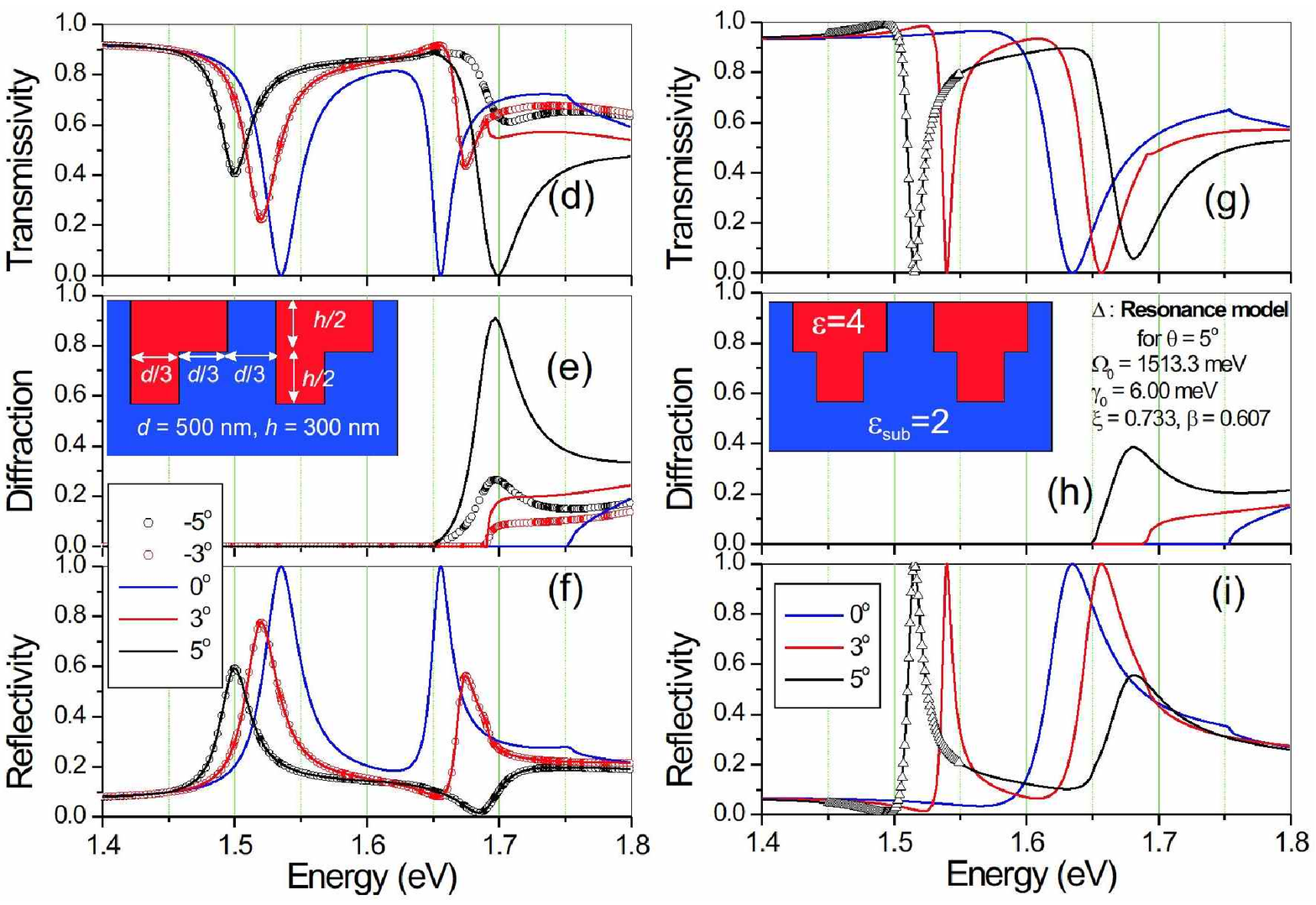}
\caption{
Calculated angular dependences of transmissivity $T = |t|^2$ (a,d,g),
reflectivity $R = |r|^2$ (c,f,i),  and total diffraction $D=1-T-R$ (b,e,h)
spectra
of the $s$-polarised light in the asymmetric (a-f) and symmetric (g-i)
1D PCS with cross-sections and parameters shown in the inserts.
For panels (d-f) and (g-i) the angles of incidence $\vartheta$  are indicated in the legends
in panels (f) and (i), respectively.
Open triangles in panels (g,i) show the resonance model Eqs.(\ref{RthetapsiSYM},\ref{TthetapsiSYM}),
the resonance parameters indicated in panel (h).
The plane of $s$-polarised light incidence is perpendicular to grooves ($\varphi = 0$).
The dips (peaks) in transmissivity (reflectivity) are quasiguided modes,
a less pronounced wrinkles (or cusps) are the diffractive Wood's anomalies.
Note that the reflectivity is symmetrical, the diffraction is asymmetrical,
and the transmissivity is symmetrical if only the diffraction is zero.
}
\label{fig2}
\end{figure*}

It follows from Eq.~(\ref{resipr}) that,
with the change of
sign of $\mathbf{k}_\parallel$ in  \textit{any} asymmetric PCS,
the reflection with  linear polarisation conservation is always symmetric,
$
    r_{ss}(\mathbf{k}_\parallel)  =  r_{ss}(-\mathbf{k}_\parallel),~
    r_{pp}(\mathbf{k}_\parallel)  =  r_{pp}(-\mathbf{k}_\parallel)
$
($r$ is the amplitude reflection coefficient).
Whereas the reflection with depolarisation from $p$ to $s$ (if exists) is asymmetric
$ r_{sp}(\mathbf{k}_\parallel)  \neq   r_{sp}(-\mathbf{k}_\parallel),$
but equals to that from $s$ to $p$,
$    r_{sp}(\mathbf{k}_\parallel)  =   r_{ps}(-\mathbf{k}_\parallel).
$
Simultaneously, the transmission (for $\mathbf{k}_\parallel$) with the conservation of linear
polarisation from the air to substrate equals always to that in the inverse
direction (for $-\mathbf{k}_\parallel$ and from the substrate to air).

To the contrary, in the case of circular polarisations,
because now the time reversal switches from $\sigma^\pm$ to $\sigma^\mp$,
the reflection
with change of polarisation is
symmetrical, $
    r_{+-}(\mathbf{k}_\parallel)  =   r_{+-}(-\mathbf{k}_\parallel). $
Simultaneously,
$  r_{++}(\mathbf{k}_\parallel)  =   r_{--}(-\mathbf{k}_\parallel),$
but the reflection with conservation of polarisation is
asymmetrical,
$ r_{++}(\mathbf{k}_\parallel)  \neq  r_{++}(-\mathbf{k}_\parallel)$. Additionally,
 $   r_{+-}(\mathbf{k}_\parallel)  \neq   r_{-+}(-\mathbf{k}_\parallel).
$

These symmetry properties of the \textit{reflection} (and \textit{transmission}
in \textit{reverse} direction) are independent of how many diffraction channels are open. They are even more
general and hold for PCS with absorptive materials, simply because any
absorption channel can be included into a more general unitary S-matrix
of the full system as an additional scattering channel.
The symmetry of transmission in reverse direction is just what was
found in Refs.~\cite{Ebbesen98,Ghaemi98,Schroeter99}. There is no contradiction here
with the different field distributions, because the reciprocity
exists between the time-reversal channels only. Whereas the full solution is not reciprocal,
due to different reflected waves in the cases of incidence from the air and the substrate.

As to the symmetry of \textit{transmission} for illumination from the \textit{same} side
of the PCS,  in case of
asymmetric PCS it holds at frequencies when
all diffraction channels are closed, and only in transparent PCS,
because the $\mathbf{k}_\parallel$ and $-\mathbf{k}_\parallel$ transmission
processes from the same side of the PCS are not reciprocal, see in Fig~\ref{fig:1a}.

We give an example of such asymmetric optical response
in Fig.~\ref{fig2}(a-f)
of a model transparent asymmetric 1D PCS,
see the schematic crossection in panel (e).
The incoming light is $s$-polarised (electric field parallel to the PCS grooves).
The calculations are done via $\mathbb{S}$-matrix method from Ref.~\cite{Tikhodeev02}
using 21 harmonics in Eq.~(\ref{kG}) with accuracy better than 10$^{-3}$.
Note that so far the diffraction is absent (for photon energy below
$\approx$1.65 - 1.75~eV, depending on the angle of incidence) the transmissivity is
symmetrical with the sign of $\mathbf{k}_\parallel$. Note also that, in agreement with the
discussion above, the reflectivity is always symmetric, see in panels (c,f).

The transmissivity dips  in Fig.~\ref{fig2} are  Wood's anomalies~\cite{Wood902,Fano41,Hessel65} due
to  quasiguided modes in PCS.
Near such resonances the reflectivity can become full in the zeroth diffraction order.
Numerical examples of such a full reflectivity can be seen in
Figs.~\ref{fig2}c,f for normal incidence in asymmetric 1D PCS,
and for {\it any} $\vartheta$  in a symmetric PCS,  Fig.~\ref{fig2}i.
The full reflectivity is achievable unless the first diffraction channel opens,
compare with panels (b,e,h).

These properties of the  resonant
reflectivity in transparent PCS can be deduced from the general properties of the
unitary scattering matrix $S_u$.
Let
$\mathrm{det}\,\mathbb{S}^{-1}(\omega,\mathbf{k}_\parallel)$ has a zero at
$\omega_0(\mathbf{k}_\parallel) =
\Omega_0(\mathbf{k}_\parallel) - i \gamma_0(\mathbf{k}_\parallel)$ in the lower half of complex
energy
plane, corresponding to the guasiguided photonic mode~\cite{Tikhodeev02}
\begin{equation} \label{Sn0}
\mathbb{S}^{-1}(\omega_0,\mathbf{k}_\parallel)|\mathbb{O}(\mathbf{k}_\parallel)\rangle = 0,
\end{equation}
where $|\mathbb{O}\rangle$ is the resonant output eigenvector in the ``large''
basis. Here we analyse the simplest case of a nondegenerate quasiguided mode.

$S_u$ has to be unitary for real $\omega$, and it  imposes  significant
restrictions on its possible form. In fact,
near the resonance one of the scattering phases (the resonant phase)
is a quickly changing function of energy,
and Eq.(\ref{SVD}) can be rewritten in a form
\begin{eqnarray} \label{SuRes}
S_u (\omega,\mathbf{k}_\parallel) =
 \sum_{j\neq 1}\mathrm{e}^{i\beta_j}|o\rangle_{j\,j}\langle \widetilde{o}| +
 \eta(\omega,\mathbf{k}_\parallel)\,|o\rangle_{1\,1} \langle \widetilde{o}|, \\
  \eta(\omega,\mathbf{k}_\parallel)  \equiv  \mathrm{e}^{i\beta_{1}} = -
 \frac{\omega-\omega^*_0(\mathbf{k}_\parallel)}{\omega-\omega_0(\mathbf{k}_\parallel)}
 = \frac{2i\gamma_0(\mathbf{k}_\parallel)}{\omega-\omega_0(\mathbf{k}_\parallel)}-1,
\end{eqnarray}
where $ |o\rangle_1 \equiv|o(\mathbf{k}_\parallel)\rangle_1$,
i.e., the resonant
``small'' outgoing vector in the orthogonal basis
corresponding to the ``large'' resonance vector $|\mathbb{O}\rangle$,
$|\tilde{o}\rangle_1 \equiv |o(-\mathbf{k}_\parallel)^*\rangle_1$,
and $|o\rangle_{j\neq 1}$ are the output basis vectors in the
subspace \textit{orthogonal} to $|o\rangle_1$.
To ensure  $\eta(\omega,\mathbf{k}_\parallel) = \eta(\omega,-\mathbf{k}_\parallel)$
(the reciprocity of the reflection near the resonance),
we have to set $\omega_0(\mathbf{k}_\parallel)=\omega_0(-\mathbf{k}_\parallel)$.
However the resonant vectors  $| o (\mathbf{k}_\parallel) \rangle_1$
and $| o (-\mathbf{k}_\parallel) \rangle_1$
can be different if there is no additional symmetry of the structure
transforming $\mathbf{k}_\parallel$ to $-\mathbf{k}_\parallel$.
Equation~(\ref{SuRes}) is the most general form of the Breit-Wigner formula (see, e.g.,
in textbook~\cite{LLQM}) for the resonant optical response in a transparent PCS.
Except $\eta$, all quantities in Eq.~(\ref{SuRes})
are slow functions of $\omega$; neglecting this $\omega$ dependence
gives a very good resonant approximation for $S_u$ in the vicinity of $\Omega_0$~\cite{NAG}.

In the zeroth  diffraction order
Eq.~(\ref{SuRes}) can be simplified further. In this case  $S_u$ is a $4\times 4$ matrix.
However, in the case of 1D PCS, and for linearly polarised light with plane of
incidence perpendicular to the grating grooves ($\varphi = 0$),
the $s$ and $p$ polarisations are decoupled.
Then, $S_u$ becomes effectively a $2\times2$ matrix.
By a proper selection of the outgoing harmonics phases, the corresponding
2D vector $|o\rangle_1$ can be made real, and
\begin{equation}\label{1dpsi}
 |o \rangle_1 = \begin{pmatrix}
 \sin \xi\\ \cos \xi
 \end{pmatrix},\,
  |o \rangle_2 = \begin{pmatrix}
\cos \xi \\ -\sin \xi
 \end{pmatrix}.
\end{equation}
The most general form of the $(2\times 2)$-dimensional
unitary scattering matrix in this
polarisation becomes
\begin{eqnarray}\nonumber
 S_u(\omega, \mathbf{k}_\|) &=&
\begin{pmatrix}
 \cos \xi \\
 -\sin \xi
\end{pmatrix}
\begin{pmatrix}
  \cos \tilde{\xi}, -\sin\tilde{\xi}
\end{pmatrix}
e^{i\beta} \\ \label{SuT_re2_psi} &+&
\begin{pmatrix}
 \sin \xi \\
 \cos \xi
\end{pmatrix}
\begin{pmatrix}
\sin \tilde{\xi},  &  \cos \tilde{\xi} \\
\end{pmatrix}
\eta(\omega,\mathbf{k}_\|),
\end{eqnarray}
where $\beta, \xi$, and $\tilde{\xi} \equiv \xi(-\mathbf{k}_\|)$ are slowly changing
with $\omega,\,\mathbf{k}_\|$
parameters of the system near the resonance.
For the coefficients of reflection $ r  \equiv (S_u)_{11}$ and transmission
$t  \equiv (S_u)_{21}$ we have, respectively
\begin{eqnarray}\label{Rthetapsi}
 r(\omega,\mathbf{k}_\|)   =  e^{i\beta} \cos \xi \cos \tilde{\xi}
  + \eta(\omega,\mathbf{k}_\|) \sin \xi \sin\tilde{\xi} .
\\
\label{Tthetapsi}
 t(\omega,\mathbf{k}_\|) = - e^{i\beta} \sin \xi \cos \tilde{\xi}
  + \eta(\omega,\mathbf{k}_\|) \cos \xi \sin\tilde{\xi} .
\end{eqnarray}
In the case of \textit{normal incidence} ($\mathbf{k}_\|=0$) for any PCS,
 we have obviously that $\xi = \tilde{\xi}$
and,  as a result,
\begin{eqnarray}\label{RthetapsiSYM}
 r(\omega,\mathbf{k}_\|) &=& e^{i\beta} \cos ^2 \xi
  + \eta(\omega,\mathbf{k}_\|) \sin ^2 \xi , \,\,\, \\\label{TthetapsiSYM}
   t(\omega,\mathbf{k}_\|) &=& [ \eta(\omega,\mathbf{k}_\|)
  - e^{i\beta} ] \cos \xi \sin\xi .
\end{eqnarray}

Obviously $t=0$ and $|r| = 1$ at the energy when $\eta(\omega) = e^{i\beta}$.
Because $\beta$ is a slow function of energy,
and $\eta$ makes a circumnutation from $\eta(\omega \ll \Omega_0 - \gamma_0) \approx -1$
through $\eta(\Omega_0) = 1$
to $\eta(\omega \gg \Omega_0 - \gamma_0) \approx -1$,
this condition always matches at some energy near the resonance, provided
$e^{i\beta}$ is not too close to -1.

The quantity
$\sin \xi \cos \xi = 1/2 \sin 2 \xi $ reaches its maximum  $1/2$ at
$\xi = \pi/4$. Thus, the only possibility to reach a full transmission $|t| = 1 $
is to have $\xi = \pi/4$ and, as seen
from Eq.~(\ref{TthetapsiSYM}), $\eta(\omega)= -e^{i\beta}$.
For example, the full transmissivity
is reached exactly at the resonance, $\omega = \Omega_0$,
in the only case  if $\beta = \pi$ and $\xi = \pi/4$ simultaneously. This is the
case of a well known Fabry-Perot resonator with
symmetrical lossless mirrors.

In asymmetric PCS,  at \textit{oblique  incidence}, as can be understood from
the inspection of Eq.(\ref{Rthetapsi}), the reflectivity still peaks (or anti-peaks)
near the resonance energy.  However, now the maximum value of $|r|$
is obviously less than 1, because $\xi \neq \tilde{\xi}$,
i.e., the resonant vectors  $|o(\mathbf{k}_\|)\rangle_1 \neq |o(-\mathbf{k}_\|)\rangle_1$
for the asymmetric PCS.

In case if  the PCS has an additional symmetry
transforming the input channels with $\mathbf{k}_\|$ to that with $-\mathbf{k}_\|$,
we have $\xi = \tilde{\xi}$. Then Eqs.~(\ref{RthetapsiSYM},\ref{TthetapsiSYM})
hold for \textit{any} angle $\vartheta$. As a consequence, the reflection
can be full near the resonance for oblique incidence,
not only for $\mathbf{k}_\| =0$ as  in the asymmetrical
gratings. This additional symmetry can be, e.g., a vertical mirror plane
(symmetrical PCS on a substrate, see insert in Fig.~\ref{fig2}e).
The numerical examples shown in Fig.~\ref{fig2}d-f fully
agree with this analysis. We would like to add here that the property of full reflectivity
is obviously destroyed by any losses, including absorption, defect and
PCS edge scattering for finite sample size.

The resonant approximation neglecting slow $\omega$ dependences in Eqs.~(\ref{Rthetapsi})-(\ref{TthetapsiSYM}) gives a very
good description of the optical response near the resonance, see an example in Fig.~\ref{fig2}g-i.
The parameters $\Omega_0, \gamma_0, \xi, \tilde{\xi}$, and $\beta$ for any PCS can be calculated
from the full scattering matrix $\mathbb{S}$~\cite{NAG}.
This resonance approximation gives, e.g., a rigorous explanation of an intuitive
model of the interference between direct and indirect pathways~\cite{Fan02}, with minimal
number of directly calculated parameters (5 for general PCS, and 4 for normal incidence or symmetric PCS),
and it is applicable for arbitrary PCS.

To conclude, using the unitarity and reciprocity properties of the scattering matrix,
we analyze theoretically the nontrivial symmetry properties and near resonance
behavior of the
optical response in photonic crystal
slabs (PCS) with asymmetric unit cell.
As a direct consequence of the reciprocity,
 the reflection with
conservation of linear polarisation is always symmetrical,
whereas that with depolarisation, if exists, is asymmetrical.
For the circularly polarised light the opposite rule holds.
As a direct consequence of the unitarity, the PCS reflectivity
peaks to unity near the quasiguided mode resonance for normal light
incidence in the absence of diffraction,
depolarisation, and losses. For the oblique incidence
the full reflectivity
in zero diffraction order is reached only in symmetrical PCS;
in asymmetrical PCS the reflectivity maximum decreases with the  angle of incidence increase.

\begin{acknowledgments}
The authors are thankful for I.~Avrutsky, L.V. Keldysh, and V.A. Sychugov for discussions.
This work was supported  in part by the Russian Foundation for
Basic Research,  Russian Ministry of Science, and Russian Academy of Sciences.
\end{acknowledgments}


\begin{thebibliography}{99}
\bibitem{endnote31}
We use conventional notations in which
$\langle \psi | = | \psi \rangle ^\protect \dag \equiv | \psi ^*\rangle ^\intercal $,
so that $\langle \psi ^* | = | \psi \rangle^\intercal$.
\bibitem{endnote32}
\mbox{$|\mathrm{out}(\mathbf{k})\rangle = S_u(\mathbf{k})\,|\mathrm{in}(\mathbf{k})\rangle \Rightarrow
|\mathrm{out}(\mathbf{k})^*\rangle = S_u^*(\mathbf{k})\,|\mathrm{in}(\mathbf{k})^*\rangle$.}
Using Eq.~(\ref{Reciprocity})~
\mbox{$\Rightarrow|\mathrm{in}(-\mathbf{k})\rangle = S_u^*(\mathbf{k})\,|\mathrm{out}(-\mathbf{k})\rangle \Rightarrow$}
\mbox{$|\mathrm{out}(-\mathbf{k})\rangle = S_u^*(\mathbf{k})^{-1}|\mathrm{in}(-\mathbf{k})\rangle$}.
\mbox{Thus,~$S_u^*(\mathbf{k})^{-1} = S_u(-\mathbf{k})$}. Using unitarity \mbox{$S_u^{-1} = (S_u^*)^\intercal$},
we arrive at Eq.~(\ref{resipr}).
\end{thebibliography}
\end{document}